# Gradient-index nanophotonics

## Igor I. Smolyaninov [1,2]*

[1]Saltenna LLC, 1751 Pinnacle Drive #600 McLean, VA 22102, USA
[2]Department of Electrical and Computer Engineering, University of Maryland, College Park, MD 20742, USA
*Corresponding author: smoly@umd.edu

**It is demonstrated that a new kind of low-loss surface electromagnetic wave may propagate along a planar surface inside a lossy medium if the medium permittivity changes continuously across such surface. Similar to surface plasmons, the wavelength of this wave may be considerably shorter than the light wavelength in free space, which may enable its applications in super-resolution microscopy and nanolithography techniques. However, unlike plasmonics-based nanophotonic devices, which are typically built using a very limited number of low loss optical materials, the newly found class of surface waves may be supported by a much broader range of lossy media. Such materials as graphite seems to be ideal in UV nanophotonics applications. On the opposite side of the electromagnetic spectrum, similar long-range surface electromagnetic radio waves are capable of propagating underwater along a sandy seabed, which may enable efficient underwater radio communication and imaging.**

**© 2020 Optical Society of America**

It is well established in the scientific literature that interfaces separating media having different electromagnetic properties may support low-loss propagation of surface electromagnetic waves. The most well-known examples of such surface waves include surface plasmon-polaritons (SPP), which propagate along a sharp interface between a good metal and a dielectric [1], and the so-called Zenneck surface wave [2], which may arguably exist at an interface between a highly lossy conductive medium (such as graphite or seawater) and a good dielectric (such as air). The wave vector $k$ of these surface waves along the interface is defined as [1,2]:

$$k = \frac{\omega}{c}\left(\frac{\varepsilon_1 \varepsilon_2}{\varepsilon_1 + \varepsilon_2}\right)^{1/2} \quad (1)$$

where $\varepsilon_1$ and $\varepsilon_2$ are the dielectric permittivities of the neighboring media. Beside the trivial case when both $\varepsilon_1$ and $\varepsilon_2$ are mostly real and positive, Eq.(1) may produce an almost purely real answer if $\text{Re}(\varepsilon_1)$ and $\text{Re}(\varepsilon_2)$ have opposite signs, $\text{Re}(\varepsilon_1+\varepsilon_2)<0$, and the imaginary parts of both dielectric permittivities are small. This case corresponds to a SPP wave at a metal-dielectric interface (note that in the $\varepsilon_1+\varepsilon_2 \rightarrow 0$ limit the $k$ vector of the wave may become very large, which is responsible for nanophotonics applications of SPP). Another possibility may be the case when $\text{Im}(\varepsilon_1)>>\text{Re}(\varepsilon_2)$, which corresponds to the Zenneck wave at an interface between a highly lossy conductive medium and a good dielectric. However, in the latter case the resulting wave vector appears to be smaller than the wave vector of regular photons in the dielectric, which leads to the "leaky" character of this surface wave: while it should be able to propagate over a perfectly smooth interface, surface imperfections must strongly scatter the Zenneck waves into photons propagating inside the dielectric.

The goal of this paper is to consider a more general situation in which the dielectric permittivity of a medium changes gradually across some real or imaginary planar surface. We will demonstrate that a low loss propagating electromagnetic wave may be sent along such a planar surface inside a lossy medium, even if the imaginary part of medium permittivity remains very high on both sides of the surface. This new surface wave solution of the macroscopic Maxwell equations appears when the interface between two media is no longer considered to be abrupt. This surprising result is applicable to any portion of the electromagnetic spectrum from the extremely low radio frequencies (ELF) up to the visible and UV ranges.

Let us consider solutions of the macroscopic Maxwell equations in a geometry in which the medium is non-magnetic ($B=H$), the dielectric permittivity of a medium is continuous, and it depends only on $z$ coordinate: $\varepsilon=\varepsilon(z)$, as illustrated in Fig.1. Under such conditions spatial variables in the Maxwell equations separate, and without the loss of generality we may assume electromagnetic mode propagation in the $x$ direction, leading to field dependencies proportional to $e^{i(kx-\omega t)}$. The macroscopic Maxwell equations may be written as

$$\vec{\nabla}\vec{D} = 0, \; \vec{\nabla}\vec{B} = 0, \; \vec{\nabla}\times\vec{E} = i\frac{\omega}{c}\vec{B}, \text{ and } \vec{\nabla}\times\vec{B} = -i\frac{\omega\varepsilon}{c}\vec{E} \quad (2)$$

leading to a wave equation

$$\vec{\nabla}\times\left(\vec{\nabla}\times\vec{E}\right) = \frac{\omega^2 \varepsilon}{c^2}\vec{E} \quad (3)$$

Since

$$\vec{\nabla}\times\left(\vec{\nabla}\times\vec{E}\right) = -\nabla^2\vec{E} + \vec{\nabla}\left(\vec{\nabla}\vec{E}\right) \quad (4)$$

and

$$\vec{\nabla}\vec{E} = -E_z \frac{\partial \varepsilon / \partial z}{\varepsilon} ,$$ (5)

after straightforward transformations we obtain

$$-\nabla^2 \vec{E} - \vec{\nabla}(E_z \frac{\partial \varepsilon}{\varepsilon \partial z}) = \frac{\varepsilon \omega^2}{c^2} \vec{E}$$ (6)

For the $E_z=0$ (TE) polarization we obtain an effective Schrodinger equation

$$-\frac{\partial^2 E_y}{\partial z^2} - \frac{\varepsilon(z)\omega^2}{c^2} E_y = -k^2 E_y$$ (7)

while for the $E_z \neq 0$ (TM) polarization the effective Schrodinger equation is

$$-\frac{\partial^2 E_z}{\partial z^2} - \frac{\partial E_z}{\partial z}\frac{\partial \ln \varepsilon}{\partial z} - \left(\frac{\varepsilon(z)\omega^2}{c^2} + \frac{\partial^2 \ln \varepsilon}{\partial z^2}\right) E_z = -k^2 E_z$$ (8)

In the latter equation the wave function may be introduced as $E_z = \psi / \varepsilon^{1/2}$, leading to

$$-\frac{\partial^2 \psi}{\partial z^2} + \left(-\frac{\varepsilon(z)\omega^2}{c^2} - \frac{1}{2}\frac{\partial^2 \varepsilon}{\varepsilon \partial z^2} + \frac{3}{4}\frac{(\partial \varepsilon / \partial z)^2}{\varepsilon^2}\right)\psi = -\frac{\partial^2 \psi}{\partial z^2} + V\psi = -k^2\psi$$ (9)

For both polarizations $-k^2$ plays the role of effective energy in the corresponding Schrodinger equations. Let us study solutions of Eqs.(7) and (9) which have a propagating wave character (Im($k$)<<Re($k$)).

In the case of TE polarized light (see Eq. (7)) the effective potential energy is $V(z) = -\frac{\varepsilon(z)\omega^2}{c^2}$, and there are no surface wave solutions. Eq.(7) only admits propagating solutions described by planar waveguide-like distributions of $\varepsilon(z)$ (see Fig.1a) in which the dielectric permittivity is positive and almost pure real. On the other hand, the TM polarized solutions of Eq. (9) can be much more interesting.

First, let us consider the SPP-like surface electromagnetic wave solutions of Eq.(9). Let us note that by making the $E_z = \psi / \varepsilon^{1/2}$, substitution in transition from Eq.(8) to Eq.(9), and assuming continuity of $\psi$, we lose the conventional SPP solution of Eq.(8) at an abrupt interface between an ideal metal and a lossless dielectric. Indeed, since $D_z$ must be continuous at such an interface, $D_z = \varepsilon^{1/2}\psi$ cannot be continuous and non-zero while $\varepsilon(z)$ remains purely real, and while it abruptly changes sign from positive to negative (if the continuity of $\psi$ is assumed at the same time). In such a case $\psi=0$ condition must be enforced, which leads to losing the SPP solution of Eq.(8). Nevertheless, it is easy to demonstrate that Eq.(9) remains quite general and reliable enough, since it allows the recovery of the SPP-like solutions in the case of a gradual interface between a real (non-ideal) metal and a dielectric (see Fig.1b). In such a case the effective potential energy $V(z)$ in Eq.(9) is dominated by the $\frac{3}{4}\frac{(\partial \varepsilon / \partial z)^2}{\varepsilon^2}$ term. This term becomes negative and strongly attractive whenever Re($\varepsilon(z)$) passes through zero:

$$V(z) \approx \frac{3}{4}\frac{(\partial \varepsilon / \partial z)^2}{\varepsilon^2} \approx -\frac{3}{4}\frac{(\partial \text{Re}(\varepsilon) / \partial z)^2}{\text{Im}(\varepsilon)^2}$$ (10)

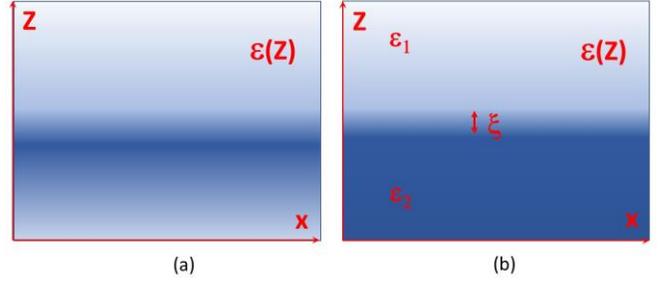

Fig. 1. Geometries of the problems of interest. The dielectric permittivity of the medium $\varepsilon$ depends only on $z$ coordinate, which is illustrated by the halftones: (a) Waveguide-like distribution of $\varepsilon(z)$. (b) Gradual interface between two media. The transition layer thickness equals $\xi$.

The deep potential well at the interface, which is described by Eq.(10) leads to the appearance of SPP-like solutions propagating along the metal-dielectric interface and having large and almost pure real $k$ vectors along the interface. Note that the Zenneck surface wave at a gradual interface between a highly lossy conductive medium and a good dielectric also arises due to the surface effective potential well described by Eq.(10).

Having recovered the conventional plasmonics-based nanophotonics within the scope of our newly developed gradient-index nanophotonics formalism, let us study what kind of other TM surface wave solutions we may obtain from Eq.(9). The effective potential energy near a gradual interface between two media shown in Fig.1b may be re-written as

$$V(z) = -\frac{4\pi^2 \varepsilon(z)}{\lambda_0^2} - \frac{1}{2}\frac{\partial^2 \varepsilon}{\varepsilon \partial z^2} + \frac{3}{4}\frac{(\partial \varepsilon / \partial z)^2}{\varepsilon^2} ,$$ (11)

where $\lambda_0$ is the free space wavelength. If $\varepsilon(z)$ of the medium changes on the spatial scale $\xi$, and this spatial scale is much smaller than $\lambda_0$, the second and the third term will dominate in Eq.(11). Moreover, if these terms are engineered (by either nanofabrication or by suitable material choice) in such a way that Im($V$)<<Re($V$), and the resulting potential well is deep enough, the wave vector of the resulting surface wave solution will be very large ($k\sim1/\xi>>2\pi/\lambda_0$) and this surface wave will have a propagating character (Im($k$)<<Re($k$)). Surprisingly enough, unlike conventional plasmonics, according to Eqs. (9, 11) a gradient-index medium which is used to support such a propagating surface wave solution does not need to be a low loss medium. For example, a medium having pure imaginary dielectric permittivity $\varepsilon(z)=i\varepsilon''(z)=i\sigma(z)/\varepsilon_0\omega$ (where $\varepsilon_0$ is the dielectric permittivity of vacuum, and the medium conductivity $\sigma(z)$ is expressed in practical SI units) will still result in Im($V$)<<Re($V$):

$$V = -\frac{i\sigma\omega}{\varepsilon_0 c^2} - \frac{1}{2}\frac{\partial^2 \sigma}{\sigma \partial z^2} + \frac{3}{4}\frac{(\partial \sigma / \partial z)^2}{\sigma^2}$$ (12)

The second and third terms in Eq. (12) are real, and they are much larger than the first term, if once again we assume that $\xi<<\lambda_0$. We should also note that the physical origins of the newly predicted surface wave solutions are different from the origins of surface plasmons. Indeed, these novel surface wave solutions may be traced back to the well-known effect of charge accumulation whenever there is a gradient of conductivity in a medium and a non-zero component of electric field parallel to it

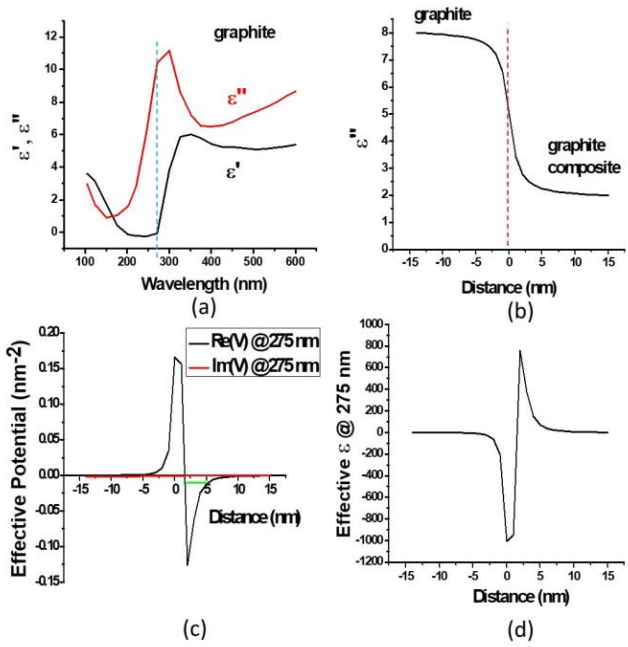

Fig. 2. (a) Real and imaginary parts of the dielectric permittivity of graphite (based on data reported in [4]). (b) Plot of an assumed $\xi=10$ nm thick planar transition layer between graphite and a graphite-based composite material. The magnitude of $\varepsilon''(z)$ is shown at $\lambda_0=275$ nm. (c) The corresponding effective potential energy (both real and imaginary parts) at the graphite interface defined by Eq.(11) (for TM light) plotted at $\lambda_0=275$ nm. The numerically obtained effective energy level is shown in green. (d) Effective $\varepsilon$ inside the surface waveguide defined by Eq.(14)

In the electrostatic case the corresponding volumetric charge density $\rho$ is obtained as

$$\rho = -\frac{\varepsilon_0 \nabla \sigma \vec{E}}{\sigma} \quad (13)$$

(see for example [3]).

Let us confirm the qualitative arguments above by detailed simulations. While the losses in a gradient optical medium do not need to be extreme to manifest the newly found surface waves, just to illustrate the point, let us consider a gradient medium based on graphite, which has almost pure imaginary dielectric permittivity in the 200-300 nm UV range [4] (see Fig.2a). Let us assume that we have engineered a planar $\xi=10$ nm thick gradual transition layer between bulk graphite and some graphite-based compound which has lower but also pure imaginary $\varepsilon$, as illustrated in Fig.2b. The resulting effective potential for TM light is shown in Fig.2c at $\lambda_0=275$ nm. If desired, based on Eq.(11) it may be represented using an "effective dielectric permittivity" distribution $\varepsilon_{eff}(z)$ (see Fig.2d) which is defined as

$$\varepsilon_{eff} = -\frac{\lambda_0^2 V(z)}{4\pi^2} \quad (14)$$

In agreement with the qualitative arguments above, we have obtained that the effective potential well near the interface is rather deep, and that Im($V$)<<Re($V$). This means that similar to any other 1D Schrodinger equation, such a potential well will always have at least one bound state. Such a bound state will give rise to at least one solution having almost pure real wave vector $k$, which corresponds to a surface mode with a long propagation length. For the effective potential well $V(z)$ shown in Fig.2c, the long-propagating-range eigenstate may be approximately determined using the virial theorem [5] as

$$k^2 \approx -\frac{\int_{-\infty}^{+\infty} \psi V(z) \psi^* dz}{2\int_{-\infty}^{+\infty} \psi \psi^* dz} \quad (15)$$

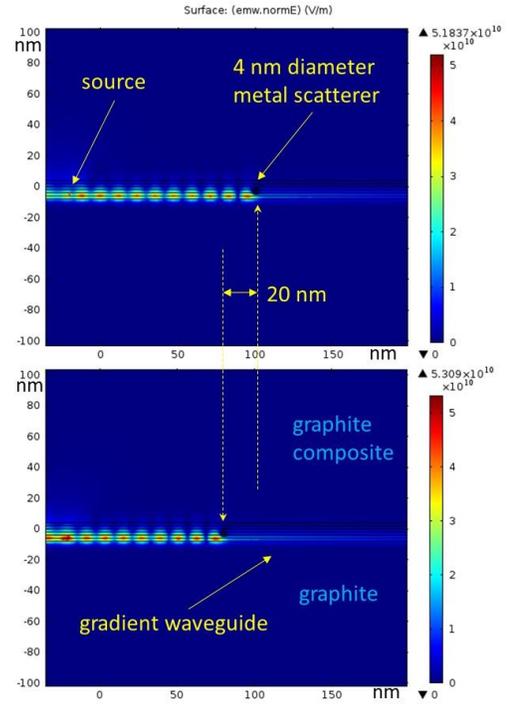

Fig. 3. Simulations of surface wave excitation and scattering in a graphite-based gradient waveguide at $\lambda_0=275$ nm. The UV light field in the waveguide is scattered by a 4 nm diameter metal nanoparticle. A 20 nm shift in nanoparticle position leads to a drastic change in the UV light distribution.

(due to almost $\sim 1/z$ functional behavior of $V(z)$ near the potential barrier located in the vicinity of the graphite surface, see Fig.2c). Alternatively, the effective Schrodinger equation may be solved numerically using the Numerov method. The numerically obtained effective energy level is shown schematically in green in Fig.2(c). The wavelength of the resulting surface wave solution is $\lambda=2\pi/k$, and $L=\text{Im}(k)^{-1}$ defines the propagation distance of the wave. Based on the numerical solution, it appears that at $\lambda_0=275$ nm the surface wave propagation distance reaches about 500 nm, which considerably exceeds the surface wave's wavelength computed numerically as $\lambda=2\pi/\text{Re}(k)=67$ nm. This clearly indicates the "propagating" character ($\lambda<<L$) of the surface wave. The newly found long-range propagating surface wave is tightly localized near the interface. Based on Eq.(9), far from the interface its electric field attenuates as $E_z \sim e^{-kz}$ away from the interface inside both media. In the particular case shown in Fig.2(c), this means that the attenuation distance away from the interface equals $l=1/k\sim 11$ nm. The field configuration in this wave is partially longitudinal, since the electric field component along the propagation direction is non-zero. Based on Eqs.(2), far from the interface $E_x \sim iE_z$, while the only nonzero component

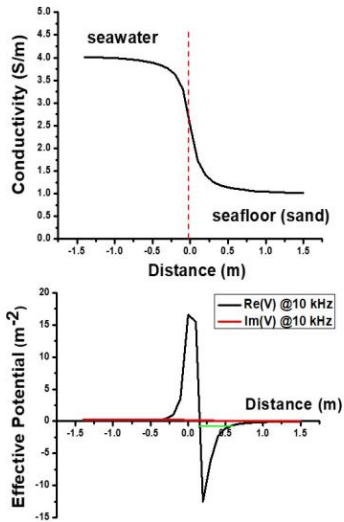

Fig.4. (a) Plot of an assumed seabed $\sigma(z)$ transition layer in which conductivity changes from $\sigma_{water}$ to $\sigma_{sand}$ within about 0.5 m. (b) The corresponding effective potential energy (both real and imaginary parts) at the water-seabed interface defined by Eq.(12) plotted for the 10 KHz band. The numerically obtained effective energy level is shown in green.

of the magnetic field is $B_y \sim -(2kc/\omega)E_z \sim -(2\lambda_0/\lambda)E_z$.

Potential applicability of this novel surface wave in super-resolution microscopy and nanolithography techniques is illustrated in Fig.3, which depicts numerical simulations of surface wave excitation and scattering in a graphite-based gradient waveguide at $\lambda_0$=275 nm. In these simulations the UV light field in the graphite-based gradient surface waveguide is scattered by a 4 nm diameter metal nanoparticle. A 20 nm shift in nanoparticle position leads to a drastic change in the UV light distribution, which is consistent with the nanometer-scale wavelength of the surface wave.

While the main emphasis of this work is directed towards nanophotonics, the obtained results have broad applicability in other applications where it is desirable to establish electromagnetic communication or to perform electromagnetic imaging through a highly lossy conductive medium. The developed theoretical model may be applicable to the imaging of underground structures and mineral deposits, radio communication along the underwater seabed and through a plasma layer around a hypersonic vehicle (such as a returning space capsule), improving propagation through electromagnetic metamaterial structures which typically suffer from high losses, bioimaging, and many other applications. The obtained results have broad applicability in any situation in which electric conductivity changes gradually across some real or imaginary interface. Such interfaces may support low-loss propagation of surface waves of charge density.

For example, based on Eq.(11) we may suggest (quite counter-intuitively) that epsilon near zero (ENZ) metamaterials [6] also become very attractive in linear and nonlinear nanophotonics applications. Despite very large light wavelengths $\sim \lambda/\varepsilon^{1/2}$ in such metamaterials, a gradient index waveguide made of ENZ metamaterials may exhibit very tight light localization (due to the novel surface wave effects described above), which in combination with nonlinear optical effects should lead to very interesting nanophotonic devices.

On the opposite side of the electromagnetic spectrum, similar long-range surface electromagnetic radio waves should be capable of propagating underwater along a sandy seabed, which may enable efficient underwater radio communication and imaging. Such a striking effect of long-distance radio wave propagation through a highly conductive medium is considered in Fig. 4 for the case of signal propagation along the underwater seabed in the 10 KHz frequency band. The case shown in Fig.4(a) assumes a sandy seabed with typical large-scale roughness of the order of 0.5 m, which determines the width $\xi$ of the $\sigma(z)$ transition layer. The sandy seabed conductivity of $\sigma$=1 S/m is assumed in these simulations based on the experimental measurements reported in [7], while the average conductivity of seawater is assumed to be $\sigma$=4 S/m. The resulting effective potential is shown in Fig.4(b) for the 10 KHz band (note that once again Im($V$)<<Re($V$)). Based on the numerical solution of the resulting Schrodinger equation (the obtained effective energy level is shown in green in Fig.3(b)) it appears that a 10 kHz radio signal propagation distance in this case reaches about $L \sim$50 m, which considerably exceeds the classical skin depth of about 3 m at 10 kHz in seawater [8]. The surface wave propagation distance is about seven times larger than its wavelength computed numerically as $\lambda = 2\pi/k$=7 m, which clearly indicates the "propagating" character ($\lambda<<L$) of the surface wave. Using a good radio receiver, which are typically capable of operating down to at least $\sim$ -100 dBm signal levels, should allow communication distances of the order of 500 m along the sandy seabed.

**Funding sources.** This work was supported in part by DARPA/AFRL Award FA8650-20-C-7027.
**Disclosures.** The author declares no conflicts of interest.